\input mssymb.tex  

\def\CL{\hbox{{$\cal L$}}}

\def\R{{\Bbb R}}
\def\C{{\Bbb C}}
\def\Z{{\Bbb Z}}

\def\vect{{\bf t}}
\def\vecu{{\bf u}}

\def\Deltam{{\Delta\kern -.65em \raisebox{.02em}{-}\kern .35em}}
\def\del{{\partial}}

\def\h{{{1\over2}}}
\def\eps{{\epsilon}}

\def\trace{{\rm Tr\, }}

\def\tens{\mathop{\otimes}}
\def\la{{\triangleright}}
\def\isom{{\cong}}

\def\Ad{{\rm Ad}}

\def\ev{{\rm ev}}
\def\coev{{\rm coev}}

\def\id{{\rm id}}

\def\Vhaj{{V\haj{\ }}}

\def\und#1{{\underline {#1}}}
\def\haj#1{{\mathaccent20 {#1}}}
\def\<{\langle}
\def\>{\rangle}

\def\o{{}_{(1)}}\def\t{{}_{(2)}}

\def\note#1{}

\def\eqn#1#2{\begin{equation}#2\label{#1}\end{equation}}

\def\align#1{\begin{eqnarray*}#1\end{eqnarray*}}

\documentstyle[11pt, epsf]{article}
\newtheorem{lemma}{Lemma}[section]
\newtheorem{propos}[lemma]{Proposition}

\newtheorem{theorem}[lemma]{Theorem}

\begin{document}\baselineskip 22pt

{\ }\hskip 1in DAMTP/93-62, to appear in Proc. XXII DGM, Ixtapa,
Mexico, Sept. 1993 \vspace{.5in}

\begin{center} {\Large LIE ALGEBRAS AND BRAIDED GEOMETRY}
\\ \baselineskip 13pt{\ } {\ }\\ S. Majid\footnote{Royal Society
University Research Fellow and Fellow of Pembroke College,
Cambridge}\\ {\ }\\ Department of Applied Mathematics \& Theoretical
Physics\\ University of Cambridge, Cambridge CB3 9EW, U.K.
\end{center}

\begin{center}
November 1993\end{center}
\vspace{10pt}
\begin{quote}\baselineskip 13pt
\noindent{\bf ABSTRACT}
We show that every Lie algebra or superLie algebra has a canonical braiding on
it, and that in terms of this its enveloping algebra appears as a flat space
with braided-commuting coordinate functions. This also gives a new point of
view about $q$-Minkowski space which arises in a similar way as the enveloping
algebra of the braided Lie algebra $gl_{2,q}$. Our point of view fixes the
signature of the metric on $q$-Minkowski space and hence also of ordinary
Minkowski space at $q=1$. We also describe an abstract construction for
left-invariant integration on any braided group.
\end{quote}
\baselineskip 22pt

\section{Introduction}

Braided geometry is a generalisation of ordinary geometry based on the idea of
{\em braid statistics} between independent systems
\cite{Ma:bg}\cite{Ma:exa}\cite{Ma:poi}\cite{Ma:lin}\cite{Ma:fre}\cite{Ma:add}.
This includes as a special case the ideas of supergeometry but with the
supertransposition $\Psi=\pm1$ there replaced by a more general braiding where
$\Psi^2\ne \id$. Braided differentiation and integration on braided vector
spaces, braided groups and braided Lie algebras are all known. Braided
manifolds and braided Yang-Mills theory are in the pipeline. The main
conclusion is that many constructions familiar in usual or supergeometry can be
generalised to the braided case. Moreover, many constructions which are more
commonly associated with quantum groups and the theory of $q$-deformations are
more properly understood in these terms. There is a review article for
physicists\cite{Ma:introp} as well as an introductory conference
proceedings\cite{Ma:carib}.

Here we would like to use some of this braided geometry to explore a basic
conceptual problem that arises in quantum physics. The problem is that we think
of a quantum algebra of observables on the one hand as a noncommutative version
of the algebra of functions on phase space, or on the other hand as generated
by the algebra of functions on configuration space and by the enveloping
algebra $U(g)$ for $g$ a generalised momentum symmetry. These points of view
are contradictory unless it happens that we can view $U(g)$ as like the algebra
of functions on some space, the momentum part of phase space.

We will see in Section~2 that for any Lie algebra $g$, one can indeed view
$U(g)$ as the algebra of functions on a braided version of $\R^n$. So the
non-commutativity of this algebra, which we normally associate with
differential operators and quantisation, can be thought of equally well as
statistical non-commutativity like that of Grassmann variables, albeit with a
braiding $\Psi$ rather than $\pm 1$.
We call this phenomenon in which a Lie algebra or enveloping algebra of
operators is thought of instead as the coordinate functions of some space, a
{\em quantum--geometry transformation}. The very simplest example is
$U(\R^n)=\C[x_1,x_2,\cdots ,x_n]$ where the enveloping algebra of an Abelian
Lie algebra is thought of instead as polynomials in some bosonic position
coordinates $x_i$. This is the idea behind Fourier transforms and our
quantum-geometry transformation is a generalisation of this.

In fact, we have already explored this idea in the context of quantum groups in
\cite{Ma:pla}\cite{Ma:the}, where it is related to Hopf algebra duality. We
proposed the ability to make this transformation, which reverses the role of
quantum and gravitational physics, as a guiding principle for physics at the
Planck scale. Now we want to touch upon these same ideas in the context of Lie
algebras and their generalisations. In fact, the above remarks apply just as
well to superLie algebras and the braided Lie algebras introduced in
\cite{Ma:lie}. In each case the enveloping algebra can be viewed instead as a
braided version of flat space. We develop this in Section~3. It provides a new
way to think about the definition of Lie algebras and braided Lie algebras.

In Section~4 we focus on the example of the braided Lie algebra $gl_{2,q}$. Its
enveloping algebra recovers a natural definition of $q$-Minkowski space. The
quantum-geometry transformation takes the subalgebra $U_q(su_2)$ to the
mass-shell in $q$-Minkowski space. The signature of the metric is also fixed as
a deformation of the Lorentzian one in this approach. As far as I know, the
Euclidean metric on $\R^4$ cannot be deformed in the same way. Thus the ability
to $q$-deform spacetime provides in this way a kind of regularity principle
that physics should not be too much an artifact of setting $q=1$. This is in
addition to the more usual motivation for $q$-deformation in terms of
regularising infinities in physics\cite{Ma:reg} and quantum corrections to
geometry.

It is hoped that this note will serve as an introduction for physicists to
braided geometry and to some of its motivation. The Appendix demonstrates some
of the mathematical techniques behind braided groups and braided geometry. We
give a self-contained account of braided integration. This provides in
principle the integration on many $q$-deformed spaces.

\subsection*{Acknowledgements} I want to thank the organisers and all the staff
for a very enjoyable and memorable conference in Ixtapa.

\section{Canonical Braiding on any Lie Algebra}

A braiding on a vector space $V$ is, by definition, a map $\Psi:V\tens V\to
V\tens V$ such that
\eqn{QYBE}{\Psi_{23}\circ\Psi_{12}\circ\Psi_{23}
=\Psi_{12}\circ\Psi_{23}\circ\Psi_{12},\qquad {\rm i.e.}\qquad
\mbox{\epsfbox{YBE.eps}}}
where the suffices refer to the copy of $V$ in $V\tens V\tens V$. If one writes
$\Psi=\mbox{\epsfbox{braid.eps}}$ then this equation expresses that the two
sides are topologically the same braid as shown.

The simplest example is when $V$ is $\Z_2$-graded and $\Psi(v\tens w)=(-1)^{|v|
|w|}w\tens v$ as in supersymmetry. Of course, in this example the exchange law
is not truly braided since $\Psi^2=\id$.

\begin{propos} Let $V=\C\oplus g$ and define the linear map
\[ \Psi(1\tens 1)=1\tens 1,\quad \Psi(1\tens \xi)=\xi\tens 1,\quad
\Psi(\xi\tens 1)=1\tens\xi\]
\[ \Psi(\xi\tens\eta)=\eta\tens\xi+[\xi,\eta]\tens 1,\qquad\qquad\quad \forall
\xi,\eta\in g.\]
Then $\Psi$ is a braiding {\em iff} $[\ ,\ ]:g\tens g\to g$ obeys the Jacobi
identity. It has minimal polynomial
\eqn{minpoly}{(\Psi^2-\id)(\Psi+\id)=0}
 {\em iff} $[\ ,\ ]$ is non-zero and antisymmetric.
\end{propos}

This is an elementary computation. It says that the definition of a Lie algebra
is mathematically completely equivalent to looking for a braiding of a certain
form. We will use this principle to give a new point of view on the definition
of a braided Lie algebra in the next section.

Now in the theory of supergeometry, the simplest examples of superspaces are
supercommutative superalgebras. Thus for $\R^{n|m}$ some of the variables (the
bosonic ones) commute and some (the Grassmann ones) anticommute etc. So the
algebra is not commutative in the ordinary sense, but it is commutative in the
super sense
\eqn{bracom}{\cdot\circ\Psi=\cdot}
where $\Psi$ is included. Likewise, the universal enveloping algebra $U(g)$ for
non-trivial Lie algebra $g$ is of course not commutative.

\begin{propos} The braiding in Proposition~2.1 extends to a braiding
$\Psi:U(g)\tens U(g)\to U(g)\tens U(g)$ and $U(g)$ is indeed braided
commutative in the sense of (\ref{bracom}).
\end{propos}

The proof of this is easy enough at degree 2 for there it says that
$\cdot\circ\Psi(\xi\tens \eta)=\eta\xi+[\xi,\eta]$ is to equal $\xi\eta$, which
is the defining relation of the universal enveloping algebra.  So imposing the
relations of braided-commutativity at order two and for the above braiding is
mathematically equivalent to the usual definition of the enveloping algebra.
The easiest way to prove the result to all orders is to prove it in complete
generality for any Hopf algebra, of which $U(g)$ is an example with coproduct
$\Delta \xi=\xi\tens 1+1\tens \xi$. If $H$ is a Hopf algebra then
\eqn{brahopf}{ \Psi(h\tens g)=\sum \Ad_{h\o}(g)\tens h\t,\qquad \Ad_h(g)=\sum
h\o g S h\t}
for all $h,g\in H$ is a braiding, and $H$ is braided commutative with respect
to it in the sense of (\ref{bracom}). Here $\Delta h=\sum h\o\tens h\t$ is the
coproduct of the Hopf algebra and $S$ is its antipode or `inverse' operation.

We see that every enveloping algebra can be regarded as the algebra of
functions on some braided space, and every quantum group too, with a suitable
choice of braiding. This change in point of view in which an enveloping algebra
gets regarded as a function algebra of some type is what we have called a
quantum-geometry transformation in the introduction. Viewing a Lie algebra
enveloping algebra in this way is significant for it means that the whole
machinery of braided spaces and braided geometry\cite{Ma:introp}, such as
braided differential operators, etc can be applied. We will compute how one or
two of these constructions look for our enveloping algebra.

In particular, given a braided algebra $B$ one has the braided tensor product
$B\und\tens B$ between two copies\cite{Ma:exa}. This is an algebra in which the
two copies do not commute but rather enjoy braid statistics. The product rule
is
\eqn{bratens}{ (a\tens b)(c\tens d)=a\Psi(b\tens c)d}
where we braid $b$ past $c$ and then multiply up. This is like the supertensor
product of superalgebras. Here is an example of what this is good for:

\begin{propos} Let $B=U(g)$ be regarded as a braided space as above. There is
an algebra homomorphism $\Deltam:B\to B\und\tens B$ given by $\Deltam\xi=1\tens
\xi-\xi\tens 1$.
\end{propos}

Just as the usual coproduct corresponds to addition (e.g. of angular momentum),
so this map corresponds to subtraction. In a dynamical context the usual
addition provides a realisation of the centre of mass system in the tensor
product of two systems, whereas the above map is more like the realisation of
the reduced mass system in the (braided) tensor product. It has properties that
one would expect for subtraction in relation to the addition. It also
generalises  to any quantum group with $\Deltam(h)=Sh\o\tens h\t$.

Now we come to a matrix version of the above results, in which we shall do a
few concrete calculations. If we choose a basis $V=\{x_\mu\}$ and write
\[\Psi(x_\mu\tens x_\nu)=x_\beta\tens x_\alpha {\bf
R}^\alpha{}_\mu{}^\beta{}_\nu\]
then the requirement for $\Psi$ to be a braiding is the celebrated Quantum
Yang-Baxter Equation (QYBE) for ${\bf R}$.

Let $g=\{x_i\}$ for $i=1,2,\cdots ,n-1$ and let $x_0=1$ so that $V=\C\oplus g$.
We use greek indices when the whole range $0,\cdots,n-1$ is intended. Then the
content of Proposition~2.1 is that
\eqn{g-R}{{\bf R}=\pmatrix{1&0&0&0\cr 0&I&0 & c\cr 0&0&I&0\cr 0&0&0&I}}
where $I$ are identity matrices and $c^i{}_{jk}$ are the structure constants of
$g$. The basis for $V\tens V$ used here is $\{x_0\tens x_0,x_0\tens
x_j,x_i\tens x_0,x_i\tens x_j\}$. Explicitly,
\[ {\bf R}^0{}_i{}^k{}_j=c^k{}_{ij},\ {\bf
R}^i{}_j{}^k{}_l=\delta^i{}_j\delta^k{}_l,\  {\bf
R}^0{}_0{}^i{}_j=\delta^i{}_j={\bf R}^i{}_j{}^0{}_0,\ {\bf R}^0{}_0{}^0{}_0=1\]
and zero for the rest. This obeys the QYBE {\em iff} $c$ obeys the Jacobi
identity.

Next, given any R-matrix, the corresponding braided space $\Vhaj(R)$ is the
algebra with $x_i$ and 1 as generators and relations
\[ x_\mu x_\nu=x_\beta x_\alpha {\bf R}^\alpha{}_\mu{}^\beta{}_\nu.\]
This defines a braided version of $\R^n$. Such a structure arises in many areas
in physics and is often called the Zamolodchikov or exchange algebra. Putting
in the form of our R-matrix (\ref{g-R}) we recover the commutation relations
\[ {} [\lambda,x_i]=0,\ [x_i,x_j]=\lambda x_k c^k{}_{ij}\]
so that the associated braided space is our enveloping algebra $U(g)$ in a
homogenised form where we add the central element $\lambda=x_0$ on the right
hand side. This is a concrete version of Proposition~2.2.

In the point of view of quantum or braided linear algebra\cite{Ma:lin}, this is
just one of many other constructions. If the $\{x_\mu\}$ are like a row vector,
then another algebra $V(R)$ defined by generators $1$ and $\{p^\mu\}$ and
relations
\[ {\bf R}^\mu{}_\alpha{}^\nu{}_\beta p^\beta p^\alpha=p^\mu p^\nu\]
is more like a column vector. For our R-matrix above, this comes out as
\[ [p^\mu,p^\nu]=0.\]
There is also a  notion of braided-quantum mechanics generalising the
one-dimensional case $px-qxp=\hbar$ to any R-matrix. It is generated by vector
and covector algebras and cross relations
\[ p^\mu x_\nu-x_\alpha {\bf R}^\alpha{}_\nu{}^\mu{}_\beta p^\beta=\hbar
\delta^\mu{}_\nu\]
as studied by several authors\cite{Kem:sym}\cite{Ma:fre}. See also the
contribution of A. Kempf at this conference. For our R-matrix (\ref{g-R}), this
comes out as
\[ [p^i,x_j]=\lambda c^i{}_{jk} p^k+\hbar\delta^i{}_j,\quad
[p^i,\lambda]=0,\quad [\pi,x_i]=0,\ [\pi,\lambda]=\hbar\]
where $\pi=p^0$. Some natural $xx$ and $pp$ relations in this context are with
a certain matrix $R'$ rather than $R$, for in this case (or in the free case
with no $xx$ or $pp$ relations) the general machinery in \cite{Ma:fre} says
that one can represent $p^\mu$ by braided differentials $\del\over\del x_\mu$
in analogy with usual quantum mechanics. One can likewise compute for our
R-matrix (\ref{g-R}) all the other R-matrix constructions for
quantum groups and braided groups. On the quantum group side one has for
example the usual quantum matrices $A({\bf R})$. This comes out essentially as
a matrix of $n$ copies of the homogenised Lie algebra, one for each row, and
with each copy transforming as an adjoint tensor operator with respect to the
others.

Finally, we note that all the constructions above work equally well if we begin
with a superLie algebra. Now the canonical braiding is
\[ \Psi(\xi\tens\eta)=(-1)^{|\xi| |\eta|}\eta\tens\xi+[\xi,\eta]\tens 1\]
and obeys (\ref{QYBE}) {\em iff} $[\ ,\ ]$ now obeys the superJacobi identity.
It obeys (\ref{minpoly}) {\em iff} $[\ ,\ ]$ is graded-antisymmetric. The
superenveloping algebra is once again characterised by (\ref{bracom}). More
generally, if $\Psi_0$ is any other symmetric braiding in the sense that
$\Psi_0^2=\id$ then for
\[ \Psi(\xi\tens\eta)=\Psi_0(\xi\tens\eta)+[\xi,\eta]\tens 1\]
to obey (\ref{QYBE}) and (\ref{minpoly}) recovers the obvious axioms of a
general $\Psi_0$-Lie algebra as in \cite{Gur:yan}. The corresponding matrix
picture is
\[{\bf R}=\pmatrix{1&0&0&0\cr 0&I&0 & c\cr 0&0&I&0\cr 0&0&0&R_0}\]

\section{Braided-Lie Algebras}

In this section we go beyond the super case and its obvious generalisations, to
the case when our Lie algebra is of a type where the background $\Psi_0$ is
itself truly braided. The axioms for such a braided Lie algebra have been
introduced by the author in \cite{Ma:lie} and consist of a coalgebra
$\CL,\Delta,\eps$, a braiding $\Psi_0=\mbox{\epsfbox{braid.eps}}:\CL\tens\CL\to
\CL\tens \CL$ and a map $[\ ,\ ]:\CL\tens\CL\to \CL$ such that
\[ \epsfbox{Liefrag.eps}\]
Here $\Delta:\CL\to \CL\tens\CL$ should be coassociative in an obvious sense
and $\eps:\CL\to\C$ should be a counit and obey $\eps\circ [\ ,\
]=\eps\tens\eps$. Note that an ordinary Lie algebra obeys these axioms if one
puts $[1,\xi]=\xi$, $[\xi,1]=0$ and
\[\CL=\C\oplus g,\quad \Delta 1=1\tens 1,\ \eps 1=1,\quad \Delta\xi=\xi\tens
1+1\tens\xi,\ \eps\xi=0.\]
So this structure $\Delta,\eps$ is implicit for an ordinary Lie algebra but we
never think about it because it has this standard form. The same is true for
superLie algebras, etc. But for examples of the truly braided type we need to
take a more general form.

\begin{theorem} Let $\CL,\Delta,\eps$ be a coalgebra and
$\Psi_0=\epsfbox{braid.eps}$ a compatible braiding. Then $[\ ,\ ]$ defines a
braided Lie algebra implies that
\[ \Psi\ =\ {{}\atop \epsfbox{Liebra.eps}}\]
is a braiding. The braided enveloping algebra $U(\CL)$ is generated by $1$ and
$\CL$ with the relations (\ref{bracom}) of braided commutativity.
\end{theorem}

The proof of this uses the same diagrammatic techniques as for braided
groups\cite{Ma:introp}. We shall see some of these techniques in action in the
Appendix. Here we content ourselves with the description of a general class of
examples from \cite{Ma:lie}. They are of matrix type where
\[\CL=\C^{n^2}=\{u^i{}_j\},\quad \Delta u^i{}_j=u^i{}_k\tens u^k{}_j,\quad \eps
u^i{}_j=\delta^i{}_j.\]
The only data we need is a matrix solution  $R\in M_n\tens M_n$ of the QYBE
which is bi-invertible. The `second inverse' here is $\widetilde R$ and is
characterised by
\[\widetilde{R}^i{}_a{}^b{}_l
R^a{}_j{}^k{}_b=\delta^i{}_j\delta^k{}_l
=R^i{}_a{}^b{}_l\widetilde{R}^a{}_j{}^k{}_b.\]
We write $I=(i_0,i_1)$ etc as multi-indices. Then\cite{Ma:skl}\cite{Ma:lie}
\[ \Psi_{0}(u_J\tens u_L)= u_K\tens u_I R_0^I{}_J{}^K{}_L,\quad
[u_I,u_J]=u_K c^K{}_{IJ}\] \[R_0^I{}_J{}^K{}_L=R^{j_0}{}_a{}^d{}_{k_0}
R^{-1}{}^a{}_{i_0}{}^{k_1}{}_b
R^{i_1}{}_c{}^b{}_{l_1} {\widetilde R}^c{}_{j_1}{}^{l_0}{}_d\]
\[c^K{}_{IJ}=\widetilde{R}^{a}{}_{i_1}{}^{j_0}{}_b
R^{-1}{}^b{}_{k_0}{}^{i_0}{}_c R^{k_1}{}_e{}^c{}_d R^d{}_{a}{}^{e}{}_{j_1} \]
is a braided Lie algebra. We changed conventions here from \cite{Ma:lie} to
lower indices for the $\{u_I\}$ in order to maintain compatibility with
Section~2. The associated canonical braiding from Theorem~3.1 is
\[ \Psi(u_J\tens u_L)=  u_K\tens u_I {\bf R}^I{}_J{}^K{}_L\]
\[ {\bf R}^I{}_J{}^K{}_L=R^{-1}{}^{d}{}_{k_0}{}^{j_0}{}_{a}
R^{k_1}{}_{b}{}^{a}{}_{i_0}R^{i_1}{}_c{}^b{}_{l_1} {\widetilde
R}^c{}_{j_1}{}^{l_0}{}_d.\]
The braided enveloping algebra $U(\CL)$ is given by taking $\vecu=\{u^i{}_j\}$
as generators and imposing $\cdot\circ\Psi=\cdot$. So this is the algebra
\eqn{B(R)}{u_Ju_L= u_Ku_I {\bf R}^I{}_J{}^K{}_L,\quad {\rm i.e.}\quad
R_{21}\vecu_1R_{12}\vecu_2= \vecu_2 R_{21} \vecu_1 R_{12}}
where the second puts two of the $R$'s to the left and uses a popular notation.

Our construction of braided Lie algebras works over the whole moduli space of
bi-invertible solutions $R$. Inside this moduli space is a subvariety of
so-called triangular solutions where $R_{21}R=1$. On this subvariety one has
$\Psi_0^2=\id$ and our braided Lie algebras are not truly braided. They reduce
in this case to the more obvious notion of $\Psi_0$-Lie algebras as at the end
of the last section after one takes a suitable scaling limit. To see this, we
parametrise $R$ in such a way that as a parameter $q\to 1$, we land on the
triangular subvariety. We also change variables to $\chi_I=u_I-\delta_I$ where
$\delta_I=\delta^{i_0}{}_{i_1}$. The braided enveloping algebra then looks like
\eqn{B(R)chi}{ \chi_J\chi_L-\chi_K\chi_I{\bf
R}^I{}_J{}^K{}_L=\chi_K\left(\delta_I {\bf
R}^I{}_J{}^K{}_L-\delta_J\delta^K{}_L\right)}
and as $q\to 1$ the right hand side vanishes. But if we rescale $\chi$ to
$\bar\chi=(q^{2}-1)^{-1}\chi$ say, then the effective structure constants for
$\bar\chi$ can have a finite limit and indeed they become those of a usual,
super,  etc. Lie algebra depending on the point on the triangular subvariety
that we are landing at. Meanwhile, the coproduct
\[ \Delta\bar\chi=\bar\chi\tens
1+1\tens\bar\chi+(q^{2}-1)\bar\chi\tens\bar\chi,\quad \eps\bar\chi=0\]
becomes our standard one. In this way, ordinary, super, etc. Lie algebras are
the semiclassical limits of braided Lie algebras as we approach the triangular
subvariety. They are therefore all unified and interpolated by our notion of
braided Lie algebras. Incidentally, this shows why the classification of all
solutions of the QYBE is such a hard problem: it includes the classification of
all Lie algebras, superLie algebras and  more generally, of braided-Lie
algebras. Usual quantum enveloping algebras also fit into this
picture\cite{Ma:lie}.

So the braided enveloping algebra in the form (\ref{B(R)chi}) looks like an
enveloping algebra but in the form (\ref{B(R)}) it looks like the coordinate
functions on a braided commutative space. This is our quantum-geometry
transformation again, in a braided form.

In fact, these quadratic algebras (\ref{B(R)}) and the matrices $R_0,{\bf R}$
were introduced by the author in \cite{Ma:exa} exactly as a braided analogue
$B(R)$ of the algebra of functions on $M_n$. They are the {\em braided
matrices} associated to $R$. We recall that the more well-known quantum
matrices $A(R)$ have a matrix of noncommuting coordinate functions forming a
bialgebra or quantum group\cite{FRT:lie}. Likewise, $B(R)$ is a
braided-bialgebra or braided group. The difference is that the matrix coproduct
above extends to an algebra homomorphism
\eqn{DeltaB(R)}{ \Delta:B(R)\to B(R)\und\tens B(R)}
provided we take for $\und\tens$ the braided tensor product algebra
(\ref{bratens}). This is like the definition of a supermatrix, but with general
braid statistics.

\section{$q$-Minkowski Space}

There are many approaches to what $q$-Minkowski space should be. Here we
describe our own approach  coming out of braided geometry\cite{Ma:mec}.
Generally speaking, our approach to $q$-deforming physics is to introduce $q$
as a parameter controlling braid statistics but with the geometry otherwise
remaining commutative. Since usual Minkowski space can be thought of as
$2\times 2$ hermitian matrices, we naturally propose that $q$-Minkowski space
should be the algebra of $2\times 2$ braided hermitian matrices. This is
broadly compatible with the pioneering approach of
\cite{CWSSW:lor}\cite{OSWZ:def}, who were motivated by the possibility of
spinors when defining their $q$-Lorentz group. On the other hand, we understand
directly the full structure of $q$-Minkowski space first and come to the
$q$-Lorentz group etc. only later as a quantum group that acts covariantly on
it.

We take the well-known R-matrix associated to the Jones knot polynomial and the
quantum plane,
\eqn{R-jones}{R=\pmatrix{q&0&0&0\cr 0&1&q-q^{-1}&0\cr 0&0&1&0\cr
0&0&0&q}}
and in this case we have the braided matrix algebra $BM_q(2)$ with generators
and relations computed in \cite{Ma:exa} as $\vecu=\pmatrix{a&b\cr c&d}$
\[ qd+q^{-1}a{\rm\ central},\  ba=q^2ab,\ ac=q^2ca,\ bc=cb+(1-q^{-2})a(d-a).\]
The braid statistics from $\Psi_0$ has $qd+q^{-1}a$ bosonic but the others
mixing among themselves. The content of the braided matrix property
(\ref{DeltaB(R)}) is that we can multiply two copies $\vecu,\vecu'$ as
\[ \pmatrix{a''&b''\cr c''&d''}=\pmatrix{a&b\cr c&d}\pmatrix{a'&b'\cr c'&d'}\]
provided we remember the corresponding braid statistics.  We also showed in
\cite{Ma:exa} that our algebra has a multiplicative braided determinant ${\rm
BDET}(\vecu)=ad-q^2cb$. It is bosonic and central.

Next, we studied $*$-structures on braided matrices in \cite{Ma:mec}. For real
$q$, we have
\[ \pmatrix{a^* & b^*\cr c^*
&d^*}=\pmatrix{a&c\cr b&d}\]
so that these matrices are naturally hermitian. One has also
\[\tau\circ(*\tens *)\circ\Delta=\Delta\circ *\]
where $\tau$ denotes ordinary transposition. This is what one would expect
since the coproduct corresponds to matrix multiplication and $(A\cdot
B)^\dagger=B\cdot A$ for ordinary hermitian matrices $A,B$. We denote the
braided matrix bialgebra $BM_q(2)$ with this $*$-structure by $BH_q(2)$, the
algebra of {\em braided hermitian matrices}. Note that the situation here is in
sharp contrast to the usual axioms of $*$-quantum groups, where hermitian
quantum matrices cannot be formulated. ${\rm BDET}$ is self-adjoint.

All of this makes this particular algebra ideally suited to serve as
$q$-Minkowski
space. So we define $q$-Minkowski space as $BH_q(2)$. The generators
\[x_0=qd+q^{-1}a,\quad x_1={b+c\over 2},\quad x_2={b-c\over 2i},\quad
x_3=d-a\]
are some natural self-adjoint spacetime coordinates while BDET becomes
\[{q^2\over(q^2+1)^2}x_0^2-q^2x_1^2-q^2 x_2^2-
{(q^4+1)q^2\over 2(q^2+1)^2}x_3^2+\left({q^2-1\over q^2+1}\right)^2{q\over 2}
x_0x_3\]
and provides a real $q$-deformed Lorentz metric.

This $q$-Minkowski space has plenty of geometry associated to it, some of which
we describe now. It is evident from the description of braided matrices
(\ref{B(R)}) that they can be viewed if we want as a 4-dimensional row vector
algebra of the same general type as the $\{x_\mu\}$ in Section~2. They
therefore transform as usual under the action of the corresponding quantum
matrices $A({\bf R})$. Thus,
\eqn{lortra}{ u_J\to  u_I\Lambda^I{}_J}
is an algebra homomorphism (we have a right comodule algebra) under the
$4\times 4$ matrix quantum group
\[  {\bf R}^I{}_A{}^K{}_B\Lambda^A{}_J\Lambda^B{}_L=\Lambda^K{}_B \Lambda^I{}_A
{\bf R}^A{}_J{}^B{}_L,\quad
\Delta\Lambda^I{}_J=\Lambda^I{}_A\tens\Lambda^A{}_J\]
This quantum group provides the basis for a $q$-Lorentz group in our picture.
It has a $*$-algebra structure
\[ \Lambda^I{}_J{}^*=\Lambda^{(i_1,i_0)}{}_{(j_1,j_0)}\]
and the coaction and coproduct are $*$-algebra homomorphisms. We have taken the
quantum group line here because it is more familiar. There is an equally good
braided Lorentz group based on $B({\bf R})$ acting in the same way as a braided
comodule algebra.

Moreover, the quantum Lorentz group here maps into the dual of the Drinfeld
quantum double\cite{Dri} with the result that our approach is indeed compatible
with other proposals based on spinors\cite{CWSSW:lor}\cite{PodWor:def}. Thus,
our $A({\bf R})$ can be realised in the quantum group $A(R)\bowtie A(R)$
introduced in \cite{Ma:mor} and generated by two copies of the $2\times 2$
quantum matrices. We take these in the form $\vect\in A(R)$ and
$\vect^\dagger\in A(R_{21})$ say, with mutual relations and $*$-structure
\[ t^i{}_a R^a{}_j{}^k{}_b t^\dagger{}^b{}_l=t^\dagger{}^k{}_b R^i{}_a{}^b{}_l
t^a{}_j, \quad t^j{}_i{}^*=t^\dagger{}^i{}_j,\qquad {\rm i.e.},\quad \vect_1
R\vect_2^\dagger=\vect_2^\dagger R\vect_1.\]
The abstract picture behind $A(R)\bowtie A(R)$ as a $*$-quantum group was found
in \cite{Ma:poi} as well as its relation to the quantum double. One should use
the inverse-transpose of the dual-quasitriangular structure found there in
Proposition~12. The realisation and the resulting $2\times 2$ matrix form of
the Lorentz transformation (\ref{lortra}) is
\[ \Lambda^I{}_J=t^\dagger{}^{j_0}{}_{i_0} t^{i_1}{}_{j_1},\quad u^i{}_j\to
u^a{}_b t^\dagger{}^i{}_a t^b{}_j,\qquad {\rm i.e.},\quad\vecu \to
\vect^\dagger\vecu\vect.\]
These constructions all work for any R-matrix of real type. For
(\ref{R-jones}), one should think of our two copies of $2\times 2$ quantum
matrices as the analogue of the complexification $SL(2,\C)$ of $SU(2)$. Then
the diagonal action $\vecu\to \vect^{-1}\vecu\vect$ when $\vect$ is unitary
defines an action of the quantum group $SU_q(2)$. This in turn is the
double-cover of rotations, which appears here as $SO_q(3)\subset SU_q(2)$, the
subHopf algebra generated by expressions quadratic in the $\vect$.

All the usual geometrical ideas likewise go though without difficulty. For
example, the mass-shell or Lorentzian sphere in $q$-Minkowski space is defined
by adding the relation
\eqn{mass-shell}{{\rm BDET}(\vecu)=1}
and is preserved under the $SO_q(3)$ action as one would expect. There are also
vector fields on $q$-Minkowski space for translation\cite{Ma:lie}, and for
Lorentz transformation from (\ref{lortra}). The action of the rotational
vectors generates the quantum group $U_q(su_2)$ as
\[ X_+\la\pmatrix{a&b\cr
c&d}=\pmatrix{-q^{3\over2}c&-q^{\h}(d-a)\cr 0 & q^{-\h}c}\to [\pmatrix{a&b\cr
c&d},\pmatrix{0&1\cr 0&0}]\]
\[ X_-\la\pmatrix{a&b\cr c&d}=\pmatrix{q^{\h}b&0\cr
q^{-\h}(d-a)& -q^{-{3\over 2}}b}\to [\pmatrix{a&b\cr c&d},\pmatrix{0&0\cr
1&0}]\]
\[ H\la\pmatrix{a&b\cr c&d}=\pmatrix{0&-2b\cr 2c& 0}\to
[\pmatrix{a&b\cr c&d},\pmatrix{1&0\cr 0&-1}]\]
where the limits are as $q\to 1$ and are as one would expect.

Another interesting feature is that this mass-shell or Lorentzian sphere forms
a braided group. This parallels the way that the Euclidean sphere in the
$2\times 2$ quantum matrices $M_q(2)$ is the quantum group $SU_q(2)$. The big
difference is the $*$-structure or signature. In fact, this is part of a
general phenomenon. Just as most familiar groups have supergoup analogues,
there is a general procedure in \cite{Ma:bg} called {\em transmutation} which
turns a quantum group into a braided group in a systematic way. The formulae at
the lowest level are
\[ u^i{}_j=t^i{}_j,\ u^i{}_j
u^k{}_l=t^a{}_bt^d{}_lR^i{}_a{}^c{}_d \widetilde R^b{}_j{}^k{}_c ,\quad {\rm
i.e.},\quad \vecu=\vect,\ \vecu_1R\vecu_2=R\vect_1\vect_2\]
etc. and come out of category theory. We also gave a direct quantum groups
point of view to them in \cite{Ma:skl}. Finally we found in \cite{Ma:mec} that
this transmutation from quantum geometry to braided geometry also has the
side-effect in general of taking us from the unitary picture (our sphere in
Euclidean space) to the hermitian picture (our Lorentzian sphere). This is the
abstract reason why only braided matrices and not quantum matrices can serve in
the $q$-deformed picture if we want the Lorentzian signature. One does not see
this constraint at $q=1$.

More recently, U. Meyer in \cite{Mey:new} has found an addition law for
$q$-Minkowski space by introducing a new braiding suitable for the coaddition
$\Delta \vecu=\vecu\tens 1+1\tens \vecu$. The R-matrix for this braiding is
different from ${\bf R}$ above and provides for a better $q$-Lorentz group with
the quantum double appearing as its double cover. The addition law also
provides for braided differential calculus according to the framework of
\cite{Ma:fre} and, in principle, a translation-invariant integration as we
shall see in the Appendix below.

This completes our introduction to the braided geometry of $q$-Minkowski space.
On the other hand, we have seen in the last section that these braided
hermitian matrices are also the braided enveloping algebra of the braided Lie
algebra associated to our R-matrix. In our case this is the 4-dimensional
braided Lie algebra $gl_{2,q}$. It has basis $h,x_+,x_-,\gamma$ with
braided-Lie bracket
\align{&&[h,x_+]=(q^{-2}+1)q^{-2}x_+=-q^{-2}[x_+,h]\\
&&[h,x_-]=-(q^{-2}+1)x_-=-q^{2}[x_-,h]\\
&&[x_+,x_-]= q^{-2}h=-[x_-,x_+]\\
&&[h,h]=(q^{-4}-1)h,
\quad [\gamma,\cases{h\cr x_+\cr x_-}]=(1-q^{-4})\cases{h\cr x_+\cr x_-}}
with zero for the others. We see that as $q\to 1$ the $\gamma$ mode decouples
and we have the Lie algebra $su_2\oplus u(1)$, but for $q\ne 1$ these are
unified. There is also a braided Killing form\cite{Ma:lie} which is
non-degenerate as long as
$q\ne 1$. So $gl_{2,q}$ is an interesting braided-Lie algebra with potential
applications in physics, such as in the unification of electroweak interactions
in $q$-deformed Yang-Mills theory\cite{BrzMa:gau} with this as the gauge
symmetry. Its $su_2$ part can also serve as differential operators of orbital
angular momentum etc., along usual lines.

The quantum-geometry transformation thus connects these two conceptually quite
distinct structures. Explicitly, it is
\[ \pmatrix{h\cr x_+\cr x_-\cr\gamma}=(q^2-1)^{-1}\pmatrix{a-d\cr c \cr b\cr
q^{-2}a+d-(q^{-2}+1)}\]
and gives an isomorphism $U(gl_{2,q})\isom BH_q(2)$. So, provided $q\ne 1$
there is only one braided group in the picture. From one point of view it is
the algebra of functions on $q$-Minkowski space. From another point of view it
is the enveloping algebra of a braided Lie algebra.
But what we see at $q=1$ is two structures, depending on how we take the limit.
If we work with $a,b,c,d$ then in the limit the algebra is the commutative
algebra of functions on usual Minkowski space. If we work with
$h,x_+,x_-,\gamma$ then the limit is the highly non-commutative enveloping
algebra $U(su_2\oplus u(1))$.

The quantum-geometry transform here is valid for $q\ne 1$ and maps Lie algebras
 and their properties to geometry. For example, what from the geometrical point
of view is the mass-shell constraint (\ref{mass-shell}) in $q$-Minkowski space,
comes out from the Lie algebra or differential operator point of view as the
quantum enveloping algebra $U_q(su_2)$. Explicitly, the quantum-geometry
transform at this level becomes
\[ \pmatrix{a&b\cr c&d}=\pmatrix{q^{H}& q^{-\h}(q-q^{-1})q^{H\over 2}X_-\cr
q^{-\h}(q-q^{-1})X_+q^{H\over 2}& q^{-H}+q^{-1}(q-q^{-1})^2X_+X_-}.\]
This follows from some known results in the theory of quantum
groups\cite{FRT:lie}\cite{ResSem:cen} by putting $\vecu=l^+Sl^-$. This
connection with quantum groups is explained in full detail in \cite{Ma:skl}, to
which we refer the reader.

Likewise, what from the geometrical point of view is the time direction $x_0$
appears from the Lie algebra point of view as giving the $u(1)$ mode $\gamma$
which could appear in a gauge theory or which, for example, acts via $[\ ,\ ]$
on $q$-Minkowski space by scaling of the space coordinates $\{x_i\}$. On the
mass-shell it appears as the quadratic Casimir. In summary,  $U(gl_{2,q})$ is
both a braided enveloping algebra, such as an internal symmetry or an algebra
of differential operators acting on $q$-Minkowski space, and can be identified
with $q$-Minkowski space itself. Only remnants of this unification are visible
when $q=1$. We have seen also that the ability to develop the $q$-deformed
picture forces us from Euclidean space to Minkowski space.

We have not had room here to describe many other features of quantum and
braided geometry. Notably, in \cite{BrzMa:gau} we introduced the theory of
quantum group principal bundles and connections (gauge fields), including the
example of a Dirac monopole on a $q$-sphere. Some of this machinery can be
applied to $q$-Minkowski space. In short, a systematic $q$-deformed picture of
the main ingredients of physics is emerging, as well as some unusual phenomena
that are not very evident at the special point $q=1$.

\section*{Appendix. Braided Integration}

In this appendix we introduce the reader to some of the mathematical techniques
of braided geometry
by deriving here a formula for invariant integration. This is a problem that is
of current interest and which was posed a couple of times at the conference.
Since quantum planes, $q$-Minkowski space and many other $q$-deformed algebras
are in fact braided groups, we can apply the general theory of braided groups.
There are still some difficulties in interpreting and computing the formula for
integration, which we offer as a challenge for the interested reader.

Our main goal is to demonstrate some diagrammatic techniques as used for the
basic properties of braided groups in \cite{Ma:introp}. We refer there for full
details of the methods and notation. As well as the result here, one can also
prove Theorem~3.1 and the braided version of (\ref{brahopf}) using the same
techniques.

Briefly, let us recall that a braided algebra $B$ is an algebra with a braiding
$\Psi=\epsfbox{braid.eps}$ mapping $B\tens B\to B\tens B$. There should also be
a unit element, which we view as a map $\eta:\C\to B$. The algebra, and indeed
all our maps, should be compatible with the braiding in an obvious way. We view
it as like functions on a braided space. A braided group is such a braided
algebra equipped also with a coproduct $\Delta:B\to B\und\tens B$ and counit
$\eps:B\to \C$. This is like the definition of a quantum group with the key
difference that $B\und\tens B$ is defined with braid statistics as in
(\ref{bratens}). We saw some concrete examples in the form of the braided
matrices in Sections~3 and~4. Likewise, some quantum planes are also braided
groups with coaddition\cite{Ma:poi}. We are using the term `braided group'
quite loosely here. In general, there should also be an antipode $S:B\to B$
obeying axioms like the usual ones. One can also ask for some
braided-commutativity as in \cite{Ma:exa} but we do not need this here.

Crucial for us is the diagrammatic notation in which
$\Delta=\epsfbox{deltafrag.eps}$ and $\cdot=\epsfbox{prodfrag.eps}$. We also
suppose that our braided group has a dual $B^*$ and denote the evaluation map
$\ev:B^*\tens B\to \C$ and coevaluation map $\coev:\C\to B\tens B^*$ by
$\ev=\epsfbox{cup.eps}$ and $\coev=\epsfbox{cap.eps}$. In concrete terms, $\ev$
is usual evaluation and $\coev(\lambda)=\lambda\sum e_a\tens f^a$ for a basis
$\{e_a\}$ and dual basis $\{f^a\}$.

Our goal is to find a map $\int: B\to \C$ which assigns to a `function' in $B$
a number, and which is translation invariant under the group law. Classically
this means $\int b(h(\ ))=\int b$ for all $h$ in our group. We find
correspondingly
\[\int=\und{\trace}L\circ S^2={{}\atop\epsfbox{intfrag.eps}}\qquad
(\id\tens\int)\Delta=1\int\]
where the first is our definition of $\int$ and the second is its
translation-invariance property. Here  $\und{\trace}$ is the braided trace as
in \cite{Ma:lie} and $L$ is left multiplication, which gives the diagrammatic
form shown.

A similar formula applies for ordinary quantum groups, and we will use a
similar strategy of proof. We note that braided integrals have also been
studied in \cite{Lyu:tan} but our proof will be different. Our first step in
the proof is a lemma. We assume that $S$ is invertible, then
\[ \epsfbox{intlem.eps}\]
where the first equality is the property that $\Delta$ is an algebra
homomorphism to the braided tensor product algebra $B\und\tens B$. The second
equality uses associativity and coassociativity of the product and coproduct.
The last equality then cancels the inverse-antipode as explained in
\cite{Ma:introp}. Then
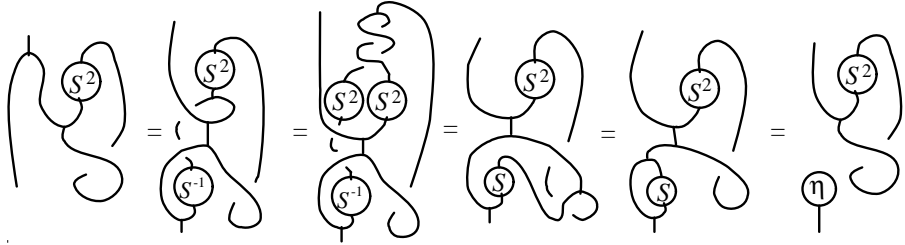
\[ \epsfbox{intproof.eps}\]
where the first equality is our lemma and the second uses that $S$ is a braided
antialgebra homomorphism. Now pick up the coproduct at the top of the third
expression and push it down and to the left (not changing the topology), giving
the fourth expression. Now we use coassociativity and cancel the antipode loop.
We obtain the desired left-invariance of the integral.

Thus we have a nice formula for the invariant integral on a braided group. The
braided trace plays the role of `averaging'. The formula should, however, be
viewed with care because it could easily happen that it gives identically zero
or infinity and may well require a renormalisation to get a finite answer. To
see the nature of this problem, let $G$ be an ordinary finite group and take a
basis of delta-functions $\{\delta_g\}$. The dual basis is the the set of group
elements themselves. Then the formula gives
\[\int b=\sum_{g}\<g,b\delta_g\>=\sum_g b(g)\delta_g(g).\]
In the continuous case this gives $\delta(0)$ times the usual integral. One can
evaluate the trace in any convenient basis. It would be interesting to find a
suitable basis in the case of the quantum plane or $q$-Minkowski space and
likewise evaluate this integral. This is a direction for further work.


\end{document}